\documentclass[conference]{IEEEtran}
\IEEEoverridecommandlockouts
\usepackage{cite}
\usepackage{amsmath,amssymb,amsfonts}
\usepackage{algorithmic}
\usepackage{graphicx}
\usepackage{textcomp}
\usepackage{xcolor}
\usepackage{amsmath}
\usepackage{booktabs,hyperref,cleveref,multirow}

\def\BibTeX{{\rm B\kern-.05em{\sc i\kern-.025em b}\kern-.08em
    T\kern-.1667em\lower.7ex\hbox{E}\kern-.125emX}}
\begin{document}

\newcommand{\gw}[1]{{\color{blue} #1}}

% \title{Conference Paper Title*\\
% {\footnotesize \textsuperscript{*}Note: Sub-titles are not captured for https://ieeexplore.ieee.org  and
\title{Perceptually Aligning Representations of Music \\ via Noise-Augmented Autoencoders
%{\footnotesize \textsuperscript{*}Note: Sub-titles are not captured for https://ieeexplore.ieee.org  and
\thanks{\textsuperscript{$^{\star}$}This research is supported by the European Research Council ERC under grant 101019375 (\textit{“Whither Music?”}) [MRB, GW].}
}

% \name{Mathias Rose Bjare$^{\star}$, \;
% Giorgia Cantisani$^{\dagger}$, \;
% Marco Pasini$^{\ddagger}$, \; 
% Stefan Lattner$^{\S}$ and \;
% Gerhard Widmer$^{\star}$
% % \thanks{ 
% % Supported by the European Research Council ERC under grants 101019375 (\textit{“Whither Music?”}) [MRB, GW].
% % %and 787836 (\textit{NEUME}) [GC]; FrontCog grant ANR-17-EURE-0017 [GC];
% % %and by a joint research project between SONY CSL and JKU [MRB, SL, GW].
% % }
% } 
  
% \address{$^{\star}$ Johannes Kepler University, Linz, AT,   
%       $^{\dagger}$ENS, PSL University, CNRS, Paris, FR, \\
%       $^{\ddagger}$Queen Mary University, London, UK, $^{\S}$ Sony Computer Science Laboratories (CSL), Paris, FR
% }

\author{
    \IEEEauthorblockN{Mathias Rose Bjare\IEEEauthorrefmark{1}, Giorgia Cantisani\IEEEauthorrefmark{2}, Marco Pasini\IEEEauthorrefmark{3}, Stefan Lattner\IEEEauthorrefmark{4} and Gerhard Widmer\IEEEauthorrefmark{1}}
    \IEEEauthorblockA{\IEEEauthorrefmark{1}Johannes Kepler University, Linz, AT,
    \IEEEauthorrefmark{2}STMS, CNRS, IRCAM, Sorbonne Université, Paris, FR\\
    \IEEEauthorrefmark{3}Queen Mary University, London, UK,
    \IEEEauthorrefmark{4}Sony Computer Science Laboratories (CSL), Paris, FR\\
    Email: \url{mathias.bjare@jku.at}
    % Georgia Institute of Technology, Atlanta, Georgia 30
    % 332--0250\\
    }
}
% \IEEEauthorblockA{\IEEEauthorrefmark{3}
% Laboratoire des Systèmes Perceptifs \\
% ENS, PSL University, CNRS, Paris, France\\
% }

% \IEEEauthorblockA{\IEEEauthorrefmark{3}Sony Computer Science Laboratories (CSL), Paris, France}
% \IEEEauthorblockA{\IEEEauthorrefmark{4}LIT AI Lab, Linz Institute of Technology, Linz, Austria\\
% Email: \url{mathias.bjare@jku.at}
% }

\maketitle

\begin{abstract}
We argue that training autoencoders to reconstruct inputs from noised versions of their encodings, when combined with perceptually motivated losses, yields encodings that are structured according to a perceptual hierarchy. 
We demonstrate the emergence of this hierarchy by showing that, after training an audio autoencoder in this manner, perceptually salient information is captured in coarser representation structures than with conventional training. 
Furthermore, we show that such perceptual hierarchies improve latent diffusion decoding in the context of estimating pitch surprisal in music and predicting EEG-brain responses to music listening.
In both cases, our results surpass those of previous methods.
Pretrained weights are available on \url{github.com/CPJKU/pa-audioic}.
\end{abstract}

\section{Introduction}
\label{sec:introduction}
Essential aspects of music appreciation, composition, and cognition are musical self-similarity, which sets expectations about the continuation of the music being listened to, and the consequent novelty or \textit{surprisal} arising when incoming sensory input defies these expectations.
The computational estimation of perceived musical expectations and surprisal has been studied using the \textit{information content} (IC) or negative log-likelihood (NLL) of autoregressive model predictions in sequentially encoded music.
The correlation between IC and surprisal has been perceptually validated in numerous behavioral and neural studies \cite{idyom_conklin,idyom,bjare2024controlling, Bjare2024audioic}.
Previous research has mainly focused on monophonic symbolic music data \cite{idyom, hansen2014predictive, bianco2020pupil, moldwin2017statistical}, which considers only a limited set of music features (e.g., monophonic pitch and timing).
In the audio domain, \cite{Skerritt-DavisE18, skerritt2019model} has proposed estimating musical surprise using Bayesian predictive inference on sequences of audio features.
However, both approaches use a few hand-selected music features that are insensitive to much of the acoustic richness of real audio and music, thereby limiting their applicability beyond perceptual experiments with controlled music stimuli.
To overcome this and the challenge of modeling high-dimensional audio data, \cite{Bjare2024audioic} estimates musical surprisal from audio autoencoder latent sequence representations \cite{pasini2024music2latent}, which are trained to minimize a perceptually inspired reconstruction loss.
The methodology has recently been extended to more powerful autoregressive diffusion models in \cite{bjare2025diffusionsurprisal}.
Notably, IC can be computed at different stages of the diffusion process, which correspond to varying levels of ``noise'' in the data.
The authors show that for appropriately moderate noise levels, the surprisal of important musical features, such as pitch, is better estimated. 
The authors hypothesize that at these noise levels, most pitch-related information is present, while information of less relevance to pitch, such as timbre nuances, is less dominant.
Spectral analysis of ordinary diffusion forward processes reveals that all frequencies of the signal entering the process (in our case, autoencoder representations) are noised equally with a strength that increases with higher noise levels \cite{dieleman2024spectral,falck2025fourier}. 
As a result, low spectral power structures (fine structures) of the signal are indistinguishable from the noise in the mixed signal and, therefore, provide no gradient to the denoising network, at lower noise levels, than structures with high spectral power (coarse structures).
In the following, we refer to this as the spectral signal-to-noise ratio (SNR) property of diffusion noise processes.
%From the spectral signal-to-noise ratios (SNR) property of these noise processes, it follows that low spectral power structures (fine structures) of the signal are indistinguishable from the noise in the mixed signal at lower noise levels than structures that have high spectral power (coarse structures). 
%During the training process of audio autoencoders, an alignment between coarse structures in representations and perceptual features (such as pitch-like qualities) is typically not enforced explicitly.
%In other words, the underlying hypothesis of \cite{bjare2025diffusionsurprisal} is not enforced explicitly.  
The hypothesis of \cite{bjare2025diffusionsurprisal} interpreted in the light of the spectral SNR property is, therefore, that an alignment between coarse structures in representations and perceptual features (such as pitch-like qualities) exists.
However, this alignment is not explicitly enforced during autoencoder training in \cite{bjare2025diffusionsurprisal}. We hypothesize that enforcing it may improve surprise estimations.
Furthermore, such alignments might increase diffusion decoding performance in general, as diffusion models produce more accurate denoisings for coarser structures than finer structures, due to the inverted U-shaped properties of the loss and modern diffusion noise schedules \cite{DBLP:conf/nips/KarrasAAL22,DBLP:conf/icml/EsserKBEMSLLSBP24}.
See \Cref{app:background} for related work on perceptual alignment in the image-pixel domain.

In this paper, we argue that a recent autoencoder training technique \cite{yang2025detok}, which adds varying amounts of noise to the latents during training, when combined with traditional perceptual loss objectives, hierarchically aligns perceptually salient features with latent structure. Under this alignment, the most salient information is captured in the coarsest structures, while progressively finer structures encode less perceptually relevant information. We demonstrate the learning of such hierarchies by finetuning the Music2Latent \cite{pasini2024music2latent} autoencoder with noise-augmented latents and show that reconstructions from latents with varying amounts of noise preserve perceptual information better in aligned latent spaces than in unaligned ones.
Furthermore, we demonstrate the importance of perceptual latent alignment for latent diffusion decoding in the case of musical surprisal estimation.
Specifically, we train autoregressive diffusion models in the aligned space to estimate surprisal in vocal and synthetic music.
Our results show that surprisal estimation is improved by the alignment procedure, as demonstrated by higher correlations with predictions of a rigorously perceptually validated \cite{ hansen2014predictive, bianco2020pupil, moldwin2017statistical} symbolic pitch expectancy model and in terms of predicting EEG brain responses to vocal music.
The estimation also improves on the results of previous methods. 
Moreover, we find the best estimations in aligned latent spaces at intermediate noise levels, whereas in unaligned spaces this is not always the case.
%This further supports that the aligned representations contain more important perceptual information in coarse structures than unaligned representations.
This shows that the hypothesis originally put forth in \cite{bjare2025diffusionsurprisal} is better supported when using representations aligned with our proposed technique than unaligned representations.

\section{Related work}
\label{app:background}

For diffusion models operating on pixels of natural images or mel-spectrograms encodings of sound, it has been shown in \cite{dieleman2024spectral, falck2025fourier} that a hierarchical alignment between coarse/fine structures and low/high frequencies is enforced by the power-law distribution of such data \cite{van1996modelling}. 
This law states that spectral power densities decrease as a power of the frequency. 
Combined with the SNR property of diffusion processes, \cite{dieleman2024spectral} therefore argues that operating in the natural data case, diffusion process generation (or IC estimation) can be viewed as an autoregression in the frequency domain. 
Furthermore, diffusion models\cite{DBLP:conf/nips/KarrasAAL22,DBLP:conf/icml/EsserKBEMSLLSBP24} typically produce high-fidelity denoising for intermediate noise levels, due to the inverted U-shaped properties of the loss and modern noise-schedules \cite{DBLP:conf/nips/KarrasAAL22,DBLP:conf/icml/EsserKBEMSLLSBP24} and, therefore, effectively produce high-fidelity results for frequencies that are not too high.
Since human perception is more sensitive to low-frequency content than high-frequency content, \cite{dieleman2024spectral} hypothesize that the autoregressive inductive bias plays an important role in the success of diffusion models.  \cite{falck2025fourier} finds that using a noise process that removes information from all frequencies uniformly can perform equally well; however, a noise process that removes information from low frequencies and then high frequencies performs substantially worse.
This shows that the order in which data features appear in the noise process plays an important role in the (autoregressive) generation process of diffusion models. 
%Recently, \cite{yang2025detok} shows that 
However, it remains less explored how these insights transfer to latent diffusion and if imposing a certain perceptual hierarchy improves such models' performance on tasks like surprisal estimation.

\section{Latent diffusion}
Latent diffusion consists of two stages.
First, an autoencoder learns to produce highly compressed data representations.
Second, a diffusion model is trained to reproduce latent encoded data.
For the first stage, we use the consistency autoencoder (CAE) of \cite{pasini2024music2latent}, composed of the encoder--decoder pair $(E,D)$.
Given an input audio sample $x$, the encoder produces a compressed latent $z = E(x)$, and the decoder reconstructs the signal as $\hat{x} = D(z)$.
In the CAE, $D$ is a (stochastic) consistency model \cite{SongD0S23} that is conditioned on the outputs of $E$.
The model is trained via consistency training \cite{SongD0S23}, which implicitly minimizes a \textit{perceptually weighted complex spectrogram} \cite{DBLP:journals/taslp/RichterWLLG23} difference (loss function) between reconstruction and input.
% The model is trained via consistency training \cite{SongD0S23}, which implicitly minimizes a perceptually weighted complex spectrogram difference between reconstruction and input: $d(\mathcal{S}_{\alpha}(x), \mathcal{S}_{\alpha}(\hat{x}))$. Here, $\mathcal{S}_{\alpha}$ denotes an amplitude-compressed complex spectrogram transform that increases the relative contribution of low-energy spectrogram components found perceptual in  \cite{DBLP:journals/taslp/RichterWLLG23}.
More generally, modern autoencoders for latent diffusion, such as \cite{DBLP:conf/icassp/EvansPCZTP25}, typically include a perceptual loss, either in the reconstruction loss or as an additional loss \cite{yang2025detok}, and could therefore be used within our framework.

For the second stage, we train an autoregressive rectified flow model \cite{DBLP:conf/nips/LiTLDH24,pasini2024continuous,bjare2025diffusionsurprisal}: a rectified flow \cite{DBLP:conf/iclr/LiuG023,DBLP:conf/iclr/LipmanCBNL23} to generate next-step predictions conditioned on a context embedding of
past observations summarized by a transformer. 
% \cite{VaswaniSPUJGKP17}.
Formally, let $z_n=z_{n,0}$ be the $n$th element of a latent representation sequence of audio.
The \textit{forward process} noises data according to: 
\begin{equation}
z_{n,t} = (1-t) z_n + t \epsilon,
\label{eq:noiseint}
\end{equation}
where  $\epsilon \sim \mathcal{N}(0, I)$.
% \begin{equation}
% p_t(z_n^{t} | z_n^0) =  \mathcal{N}((1-t) z_n , t² ). 
% \label{eq:noiseint}
% \end{equation}
An autoregressive rectified flow model, represented by a diffusion MLP $v_\phi$ conditioned on a context summarized by a causal transformer $f_{\theta}$, learns a continuous normalizing flow \cite{DBLP:conf/nips/ChenRBD18} from the noise distribution $z_{n,1}=\epsilon$ to the distribution of sequence elements $z_{n,0}$.
The flow is learned simultaneously for all $n$ by the optimization
%conditioning $v_{\phi}$ on past elements $z_{<n}$ and optimizing
\begin{equation*}
    \min_{\phi,\theta} \frac{1}{N}\sum_{n=1}^{N}\mathbb{E}_{t,\epsilon}\left[\lVert \left(z_n - \epsilon \right) - v_{\theta}\left((1-t) z_n + t \epsilon ,  f_\phi \left(z_{<n}\right)\right)\rVert_2\right],
\end{equation*}
where $t$ is sampled from a logit-normal \cite{atchison1980logistic} distribution $\text{sigmoid}( \varepsilon)$, with $\varepsilon \sim \mathcal{N}(m,s^2)$ and $m,s=0,1$ as proposed in \cite{DBLP:conf/icml/EsserKBEMSLLSBP24}.
In this paper, instead of generating samples autoregressively, we compute IC or negative log-likelihoods of next-step predictions in a teacher-forcing manner using the instantaneous change-of-variables formula \cite{DBLP:conf/nips/ChenRBD18}, as in \cite{bjare2025diffusionsurprisal}. 
%In contrast, variational autoencoders (VAEs) avoid collapse by enforcing a probabilistic structure through the objective
% \begin{equation}
% \mathcal{L}_{\mathrm{VAE}} = \mathbb{E}_{E(z|x)}\!\left[\|x - D(z)\|^2\right] + \lambda_\mathrm{KL} D_\mathrm{KL}\!\left(E(z|x)\,\|\,p(z)\right),
% \end{equation}
% where the $D_\mathrm{KL}$ term aligns $E(z|x)$ with a prior $p(z)$.

% In contrast, modern variational autoencoders (VAEs) are trained with the loss 
% \begin{equation}
% \mathcal{L}_{\text{total}} = \mathcal{L}_{\text{MSE}} 
% + \lambda_{\text{KL}} \mathcal{L}_{\text{KL}} 
% + \lambda_{\text{percep}} \mathcal{L}_{\text{percep}} 
% + \lambda_{\text{GAN}} \mathcal{L}_{\text{GAN}} .
% \end{equation}
% where the $\mathrm{KL}$ term aligns $E(z|x)$ with a prior $p(z)$ to avoid latent space collapse.

\section{Perceptual alignment}
\cite{yang2025detok} studies noise-augmenting the traditional autoencoder reconstruction learning framework for diffusion models by 
interpolating the latents with noise similar to the noising process of rectified flows.
To ease notation, we denote by $z$ any $n$th element of the clean sequence, $z_{n,0}$.  
During training, \cite{yang2025detok} noises latents $z$  with \Cref{eq:noiseint}, where instead $\epsilon \sim \mathcal{N}(0, \gamma^2)$, 
% \begin{equation}
% z' = (1 - t) z + t \, n(\gamma)\text{, where } n(\gamma) \sim  \mathcal{N}(0, \gamma^2), \; 
% t \sim \mathcal{U}(0,1).
% \label{eq:noiseint}
% \end{equation} 
and task the autoencoder to reconstruct clean data.
Although noising the latents during autoencoder training resembles the diffusion forward noise processes, we argue that it serves a fundamentally different purpose since $z$ is learned and not frozen.
Observing the reconstruction of a single input data example, the encoder has to learn representations that, when decoded, simultaneously minimize the perceptual loss for different noise levels.
Following the spectral SNR property of diffusion noise processes, this particularly means that information related to satisfying the perceptual loss should mostly be encoded in coarse latent structures, and information with increasingly less perceptual relevance should be encoded in increasingly finer structures.
We refer to this information organization as \textit{perceptual alignment}.
In contrast, the forward noise plays no role in the latent space organization.

We propose the following modifications to the method presented in \cite{yang2025detok} that we empirically found beneficial for downstream tasks.
Using $\epsilon\sim \mathcal{N}(0,\gamma^2)$, the expected SNR is given by $\mathbb{E}[z^2]/\gamma^2$, and can be controlled by $\gamma$. 
The encoder in \cite{yang2025detok} can learn to increase the variance of $z$ to increase the expected SNR, which essentially reduces the effect of noising.
We, in contrast, fix the variance of $z$ to the variance of the noise distribution using layer normalization \cite{DBLP:journals/corr/BaKH16}, such that the expected SNR stays constant during training.
%$z_t$'s variance conditioned on $t$ is $\text{Var}(z_t|t) = (1-t)^2\text{Var}(z)+t^2\gamma^2$ and depends on $\gamma$.
%This means, for example, that fully noised data could have a much larger variance than clean data, which in extreme cases might make representation learning less effective.
We set $\gamma^2=1$ 
%such that $\text{Var}(z_t|t) = ((1-t)^2+t^2)\text{Var}(z)$ and does not depend on $\gamma$ 
and control the expected SNR by sampling $t$ from a biased logit-normal distribution.
We found $m,s=-1,1$ to work best empirically. 
Unlike in \cite{yang2025detok}, our latent noise process is the same as the rectified flow noise process used for latent diffusion, except for $t$'s distribution.
%In \Cref{app:ablation}, we show the importance of fixing the variance and provide model selection details.
\begin{table}
        \centering
\caption{Perceptual quality metrics for reconstructions of aligned latents $E,D$ and unaligned latents $D$ and $\emptyset$.}
        \label{tab:autoencoder}
\begin{tabular}{lccccc}
\toprule
\scriptsize SNR & \scriptsize $NT$ & \scriptsize V ($\uparrow$) & \scriptsize SI ($\uparrow$) & \scriptsize $F_{\text{VGG}}$ ($\downarrow$) & \scriptsize $F_{\text{CLAP}}$ ($\downarrow$) \\
\midrule
\multirow{3}{*}{$\infty$}
  & $E,D$ & 3.73 & -5.18 & 1.53 & 0.05 \\
  & $D$ & 3.73 & -4.97 & 1.58 & 0.05 \\
  & $\emptyset$ & \textbf{3.84} & \textbf{-3.84} & \textbf{1.16} & \textbf{0.04} \\
\midrule
\multirow{3}{*}{$4.0$}
  & $E,D$ & \textbf{3.48} & \textbf{-9.05} & \textbf{2.46} & \textbf{0.08} \\
  & $D$ & 3.45 & -10.31 & 2.89 & 0.09 \\
  & $\emptyset$ & 2.94 & -11.44 & 6.63 & 0.42 \\
\midrule
\multirow{3}{*}{$1.0$}
  & $E,D$ & \textbf{3.19} & \textbf{-15.73} & \textbf{3.64} & \textbf{0.17} \\
  & $D$ & 3.18 & -18.52 & 3.94 & 0.19 \\
  & $\emptyset$ & 2.53 & -18.82 & 11.15 & 0.84 \\
\midrule
\multirow{3}{*}{$0.25$}
  & $E,D$ & 2.87 & -29.78 & \textbf{5.01} & 0.38 \\
  & $D$ & \textbf{2.88} & -32.17 & 5.10 & \textbf{0.35} \\
  & $\emptyset$ & 2.22 & \textbf{-28.44} & 15.04 & 1.17 \\
\bottomrule        
\end{tabular}
\end{table}

\section{Reconstruction experiments}
\label{sec:reconstruction}
We test the efficiency of the noising technique in hierarchically aligning more perceptually important features with coarse structures using the autoencoder's reconstructions.
We finetune the publicly available Music2Latent checkpoint using the same data, architecture, and hyperparameters as described in \cite{pasini2024music2latent}, except that we use a constant consistency-step schedule with a step size fixed to the final value of the pre-trained model. 
To quantify the amount of perceptually salient information encoded in coarse structures, we create reconstructions from latents that encode features at various coarseness levels. We then check the reconstructions' correspondence to the input using the perceptually motivated reconstruction metrics ViSQOL (V) \cite{hines2015visqol,sloan2017objective,chinen2020visqol}, a MOS-like distance between two audio samples, and SI-SDR (SI), a spectrogram distance \cite{le2019sdr}, as used in \cite{pasini2024music2latent,defossez2022high}.
Since it is challenging to disentangle structures of varying coarseness in the latents explicitly, we instead use the spectral SNR property and construct latents at four different coarseness levels by encoding diverse  10s music audioclips from MusicCaps \cite{agostinelli2023musiclm} and adding noise following the same latent noising process as used for training, but setting $t$ in a way that the SNR levels are $\{\infty, 4, 1, .25 \}$ respectively.
We report the results in \Cref{tab:autoencoder} as $NT=E, D$, emphasizing that both encoder and decoder have been trained with noised latents.
At low SNR levels, it is likely that information essential to faithfully reconstructing the signal is removed, leaving the decoder to infer the input's likely form.
To measure the realism of reconstructions at low SNR levels (not necessarily following the input strictly), we additionally report the distribution metrics FAD \cite{DBLP:conf/interspeech/KilgourZRS19} score using the VGGISH ($F_{\text{VGG}}$) and the CLAP ($F_{\text{CLAP}}$)\cite{DBLP:conf/icassp/WuCZHBD23} versions.
We compare the results of the perceptually aligned autoencoder against the results of an unaligned autoencoder in two scenarios: 1) where the training-inference discrepancy is fixed by freezing the encoder and training the decoder with noised latents, reported as $NT=D$, and 2) without the correction, reported as $NT=\emptyset$.
Comparing SI for $NT=E,D$ and $NT=D$ at SNR values $<\infty$, we find that the perceptual information retained in coarse structures is always higher for the aligned representations than for the unaligned representations and similar for V.
Both V,SI are higher comparing $NT=E,D$ and $NT=\emptyset$ for $SNR<\infty$, except for $SI$ at $SNR=0.25$.
At this low SNR level, where much of the input's information has been removed, the FAD score of the aligned autoencoder is much lower.
This indicates that the stochastic decoder is inventing information to create plausible reconstructions that diverge from the input.

\section{Musical surprisal experiments}
    \begin{figure}
    % Second minipage: the table
        \centering
        \includegraphics[width=\linewidth, trim={0 0 0 0.4cm}, clip]{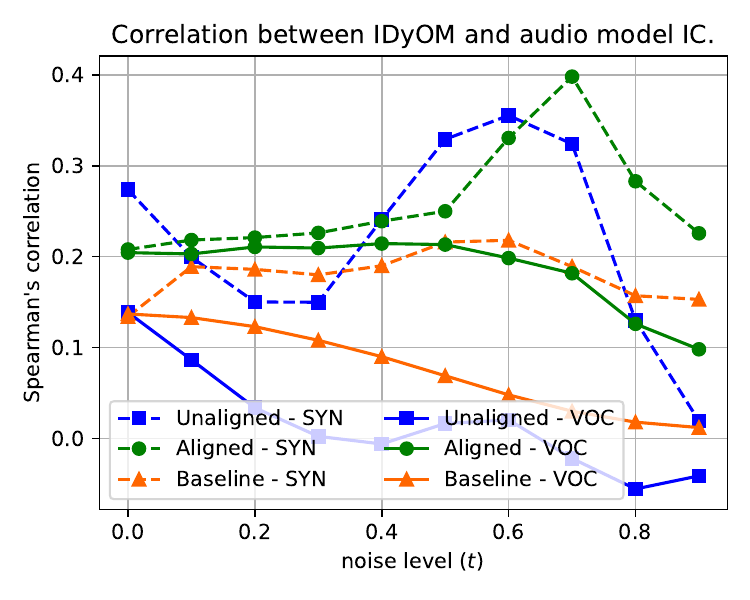} % your image
        \caption{Correlation between IC calculated on melodies using IDyOM and calculated with aligned and unaligned latents and using the baseline of \cite{bjare2025diffusionsurprisal} at different noise levels.}
        \label{tab:idyom}
    \end{figure}
\label{sec:suprisalestimation}
We are interested in whether hierarchical latent alignment improves musical surprisal estimation in the diffusion noise space continuum.
Specifically, we investigate the model's capabilities to estimate pitch surprisal and to predict EEG responses to sung music.  
For estimating pitch surprisal, we largely follow the methodology of \cite{bjare2025diffusionsurprisal} and train an autoregressive rectified flow model using the same data, model, and hyperparameters except for lowering the maximum sequence length to 3125, corresponding to $\sim$5 minutes of audio, using AdamW with base lr $10^{-4}$, 750k-step cosine schedule with 10k-step linear warmup and applying a $m,s=0,1$ logit-normal noise schedule.
For experiments involving singing voices, we finetune our model on a small private dataset of singing voices running for 36k steps, with a base learning rate of $5\times10^{-5}$ and a warmup of 12k steps.
We then compare whether alignment improves agreement between IC derived from autoregressive latent diffusion models and IC derived from IDyOM \cite{idyom}, a  perceptually validated \cite{hansen2014predictive, bianco2020pupil, moldwin2017statistical} pitch expectancy model that operates in a condensed symbolic domain. 
We conduct our experiment using a synthesized dataset (SYN) of Irish monophonic tunes, described in \cite{folkrnnsession}, and a recorded vocal dataset (VOC) along with its automatic transcription, described in \cite{cantisani2023investigating}.
We extract the IC of each note pitch in the symbolic datasets using IDyOM and pair these with IC values calculated with our models in aligned and unaligned spaces at various noise levels.
We compare the paired estimates using Spearman's rank correlation and report the results in \Cref{tab:idyom}, which are significant on a $5\%$ significance level (except for VOC unaligned correlations $t\in [0.3, 0.7]$).
We additionally report the results of \cite{bjare2025diffusionsurprisal} as Baseline.
For the aligned latent space, as the noise level decreases, the correlation increases until a maximum and then decreases.
This suggests that after reaching a certain noise level, within $[0.5, 0.7]$ for both datasets, most information relevant for pitch surprisal estimation is already present in the noised data, and adding additional information may decrease the correlation.
For the unaligned space and the baseline, this is not the case. 
The highest correlations for both datasets are found in the aligned space.
%We provide audio examples in the supplementary material for the two datasets where we reconstruct latents that are noised like in \Cref{sec:reconstruction}. These demonstrate that pitch is preserved audibly for the noise-levels where the correlations are highest, while other information e.g. related to timbre, phonemes and words are lost.      

\begin{figure}
    \centering
    \includegraphics[width=\linewidth]{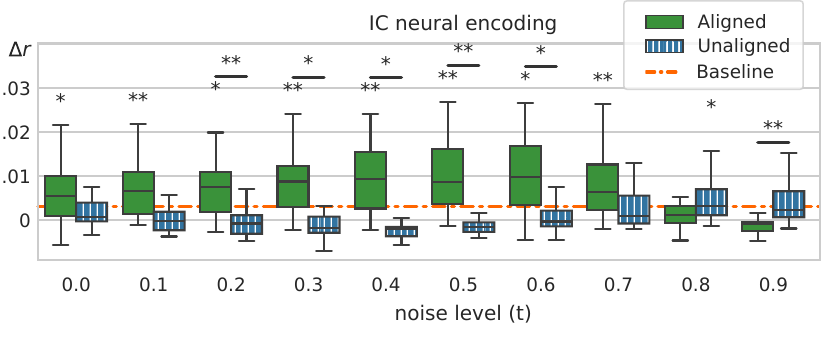}
    \includegraphics[width=\linewidth]{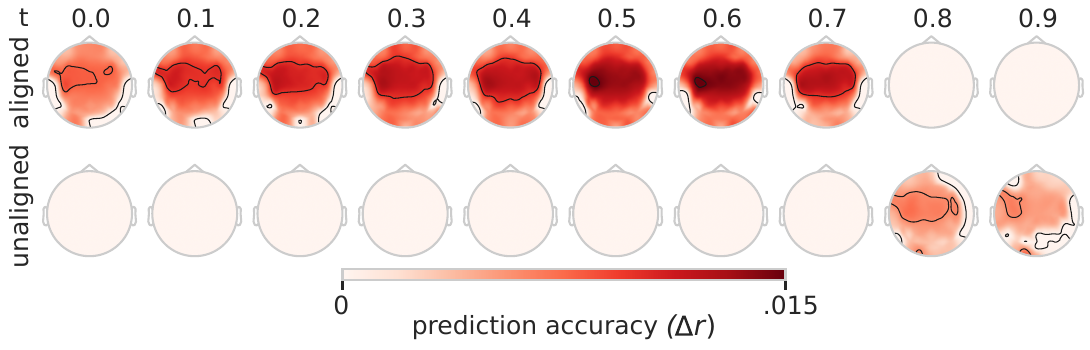}
    \caption{Cortical tracking of IC computed with aligned and unaligned latents across different noise levels. $\Delta r$ denotes the increase in prediction accuracy using a full model (IC + E) with a reduced model (only E).
Bar plots report the mean ± SE across participants (median across electrodes, average across trials).
Scalp topographies report $\Delta r$ for individual channels (only $p<0.05$ significant channels shown).}
    \label{fig:eeg1}
\end{figure}

% \begin{figure}
%     \centering
%     \includegraphics[width=.85\linewidth]{figs/subset_models_boxplot.pdf}
%     \caption{Cortical tracking of ICs computed with aligned and unaligned latents across different noise levels. $\Delta r$ denotes the increase in prediction accuracy when comparing a full model (IC + acoustic envelope) with a reduced model including only the envelope.
% Bar plots report the mean ± SE across participants (median across electrodes, average across trials).
%     \label{fig:eeg1}
% \end{figure}

To further evaluate how the proposed model correlates with human perception, we tested whether IC estimated with perceptually aligned latents predicts neural responses to music more accurately than IC estimated with unaligned latents.
To do so, we follow the methodology of \cite{Bjare2024audioic} and quantify the variance in brain responses explained by IC, \textit{i.e.,} its neural encoding.
To this end, we used Ridge regression to model EEG responses as a linear combination of two predictors: (i) IC and (ii) the acoustic envelope of the waveform (E), computed using the Hilbert transform's magnitude.
We compared the neural encoding of IC features computed
%with aligned and unaligned latents across different noise levels 
in EEG responses to the sung music of VOC (64 channels, 20 participants, 18 songs, see \cite{cantisani2023investigating,Bjare2024audioic} for details). 
%For each participant and condition, the channel-specific mappings between predictors and EEG were estimated by solving a regularized linear regression problem \cite{crosse2016multivariate}. 
Separate and independent optimal filters were estimated for each channel, participant, model, and noise levels by solving a regularized linear regression problem \cite{crosse2016multivariate}.
Non-instantaneous interactions were captured by including multiple stimulus–response time lags within a $[-100, 700]$ ms window, with an additional $50$ ms margin to avoid edge artifacts. 
Model performance was evaluated using leave-one-out cross-validation across trials, and quantified as the Pearson correlation ($r$) between the predicted and observed EEG signals at each electrode.
The significance of the IC contribution was assessed by the difference in predictive power of a full model (IC + E) and that of a reduced model (E only) reported as $\Delta r$ in \Cref{fig:eeg1}.
%and provides a measure of unique variance explained by IC.
% The difference in predictive power ($\Delta r$) provides a measure of unique variance explained by IC beyond low-level acoustics.
% % This may go in the Appendix
% Specifically, we used Ridge regression to model EEG responses as a linear combination of two predictors: IC and the acoustic envelope of the waveform. The latter served as a nuisance regressor, absorbing variance due to low-level acoustic features and trivial voiced/unvoiced responses.
Model performance was then quantified as Pearson's $r$ between predicted and actual EEG using a leave-one-out cross-validation procedure across songs.
The significance of the IC contribution was assessed by comparing the predictive power of a full model including both IC and acoustic envelope as predictors with that of a reduced model that included only the envelope. 
The IC of the aligned method produced significantly stronger cortical tracking than the IC of the unaligned method and the baseline model of  \cite{bjare2025diffusionsurprisal} across most noise levels ($t=0.2$–$0.6$; Fig.~\ref{fig:eeg1}), with the largest improvements observed around mid-level noise.
This is consistent with the highest IDyOM correlations observed in that dataset.
This advantage was consistent across participants and electrodes, as also reflected in the scalp topographies, which revealed widespread positive effects over fronto-central regions.
%At high noise ($t \geq 0.8$), the advantage of alignment disappeared, suggesting that perceptual alignment of IC is most beneficial when acoustic degradation is moderate.
These findings demonstrate that IC at moderate noise levels in perceptually aligned representations is reliably encoded in neural responses to music.
\section{Conclusion and Discussion}
Together, the two musical surprisal experiments yield consistent results, suggesting that perceptually structured latent representations may benefit tasks in music perception and latent-diffusion decoding.
This is despite the information loss caused by the noise augmentation, as suggested by the slightly lower reconstructive performance on clean data. A similar effect was found in \cite{yang2025detok}.
Future studies should investigate whether closing the gap leads to better overall results. 
In our EEG experiments, we obtained the best results at a high noise level ($t=0.6$) in aligned representations.
Interestingly, this noise level coincides with a high performance in our pitch surprisal estimation and is consistent with the result of \cite{cantisani2024neural} that predicts EEG responses only using pitch information.
The same is not true for unaligned representations. This shows that the hypothesis of \cite{bjare2025diffusionsurprisal} (see \Cref{sec:introduction}) is better supported in aligned rather than unaligned representations.
Future work should investigate musical or audio features other than pitch present in the signal at such high noise levels, as well as identify those that emerge at lower noise levels, since their presence appears to impair EEG prediction.
% however, it should also examine which features appear only at lower noise levels (as their presence negatively impacts EEG prediction). 

\clearpage
\section*{Acknowledgments}
The author thanks Jan Schlüter for helpful discussions and suggestions on the project and the manuscript.
% \gw{This research is supported by the European Research Council ERC under grant 101019375 (\textit{“Whither Music?”}) [MRB, GW].}
%and 787836 (\textit{NEUME}) [GC]; FrontCog grant ANR-17-EURE-0017 [GC];
%and by a joint research project between SONY CSL and JKU [MRB, SL, GW].

{
% \fontsize{9pt}{10pt}\selectfont
% \small
\bibliographystyle{IEEEtran}
\bibliography{strings,refs}
}
% \bibliographystyle{IEEEbib}
% \bibliography{strings,refs}
\end{document}